\documentclass[aps,prb,twocolumn,reprint,amsmath,amssymb,superscriptaddress,floatfix,footinbib,longbibliography]{revtex4-2}

\usepackage{graphicx}
\graphicspath{ {../images/} }
\usepackage{epstopdf}
\usepackage{tabularx}

\usepackage[T1]{fontenc}
\usepackage[applemac]{inputenc}
\usepackage{lmodern}
\usepackage{amsmath}
\usepackage{amssymb}
\usepackage[english]{babel}
\usepackage{natbib}
\usepackage{ae}
\usepackage{units}

\usepackage{amsmath,amssymb,natbib,bm}
\usepackage{psfrag}
\usepackage{subfigure}
\usepackage{amsthm}

\usepackage{booktabs}

\usepackage{slashed}

\usepackage[americaninductors]{circuitikz}
\usepackage{tikz}
\usetikzlibrary{arrows}
\usepackage{ulem}
\usepackage{color}
\usepackage{url}

\usepackage[colorlinks]{hyperref}
\hypersetup{%
        plainpages=true,
        breaklinks=true,
        hypertexnames=false,
        pageanchor=true,
        colorlinks=true,
        linkcolor={blue},
        citecolor={magenta},
        urlcolor={blue},
        anchorcolor={black}
      }

\usepackage{mleftright} 

\newcommand{\figref}[1]{\mbox{Fig.~\ref{#1}}}
\newcommand{\tabref}[1]{\mbox{Table~\ref{#1}}}

\newcommand{\appref}[1]{\mbox{Appendix~\ref{#1}}}
\renewcommand{\eqref}[1]{\mbox{Eq.~(\ref{#1})}}

\newcommand{\figpanel}[2]{Fig.~\hyperref[#1]{\ref*{#1}(#2)}} 
\newcommand{\figpanels}[3]{Fig.~\hyperref[#1]{\ref*{#1}(#2)-(#3)}} 
\newcommand{\figpanelNoPrefix}[2]{\hyperref[#1]{\ref*{#1}(#2)}} 
\newcommand{\figpanelsNoPrefix}[3]{\hyperref[#1]{\ref*{#1}(#2)-(#3)}} 

\newcommand{\expec}[1]{\mleft\langle #1 \mright\rangle}

\newcommand{\sz}{\hat \sigma_z}

\newcommand{\sm}{\hat \sigma_-}
\renewcommand{\sp}{\hat \sigma_+}

\newcommand{\abssq}[1]{\mleft| #1 \mright|^2}

\AtBeginDocument{%
    \newwrite\bibnotes
    \def\bibnotesext{Notes.bib}
    \immediate\openout\bibnotes=\jobname\bibnotesext
    \immediate\write\bibnotes{@CONTROL{REVTEX42Control}}
    \immediate\write\bibnotes{@CONTROL{%
    apsrev42Control,author="08",editor="1",pages="0",title="0",year="1"}}
     \if@filesw
     \immediate\write\@auxout{\string\citation{apsrev42Control}}%
    \fi
}%

\begin{document}

\title{Tuning atom-field interaction via phase shaping} 

\author{Y.-T.~Cheng}
\affiliation{Department of Physics, National Tsing Hua University, Hsinchu 30013, Taiwan}

\author{C.-H.~Chien}
\affiliation{Department of Physics, National Tsing Hua University, Hsinchu 30013, Taiwan}

\author{K.-M.~Hsieh}
\affiliation{Department of Physics, National Tsing Hua University, Hsinchu 30013, Taiwan}

\author{Y.-H.~Huang}
\affiliation{Department of Physics, National Tsing Hua University, Hsinchu 30013, Taiwan}

\author{P.~Y.~Wen}
\affiliation{Department of Physics, National Chung Cheng University, Chiayi 621301, Taiwan}

\author{W.-J.~Lin}
\affiliation{Department of Physics, National Tsing Hua University, Hsinchu 30013, Taiwan}

\author{Y.~Lu}
\affiliation{Department of Microtechnology and Nanoscience (MC2), Chalmers University of Technology, SE-412 96 Gothenburg, Sweden}

\author{F. Aziz}
\affiliation{Department of Physics, National Tsing Hua University, Hsinchu 30013, Taiwan}

\author{C.-P.~Lee}
\affiliation{Department of Physics, National Tsing Hua University, Hsinchu 30013, Taiwan}

\author{K.-T.~Lin}
\affiliation{Department of Physics and Center for Quantum Science and Engineering, National Taiwan University, Taipei 10617, Taiwan}
\affiliation{Physics Division, National Center for Theoretical Sciences, Taipei 10617, Taiwan}

\author{C.-Y.~Chen}
\affiliation{Department of Physics, National Tsing Hua University, Hsinchu 30013, Taiwan}

\author{J.~C.~Chen}
\affiliation{Department of Physics, National Tsing Hua University, Hsinchu 30013, Taiwan}
\affiliation{Center for Quantum Technology, National Tsing Hua University, Hsinchu 30013, Taiwan}

\author{C.-S.~Chuu}
\affiliation{Department of Physics, National Tsing Hua University, Hsinchu 30013, Taiwan}
\affiliation{Center for Quantum Technology, National Tsing Hua University, Hsinchu 30013, Taiwan}

\author{A.~F.~Kockum}
\affiliation{Department of Microtechnology and Nanoscience (MC2), Chalmers University of Technology, SE-412 96 Gothenburg, Sweden}

\author{G.-D.~Lin}
\affiliation{Department of Physics and Center for Quantum Science and Engineering, National Taiwan University, Taipei 10617, Taiwan}
\affiliation{Physics Division, National Center for Theoretical Sciences, Taipei 10617, Taiwan}
\affiliation{Trapped-Ion Quantum Computing Laboratory, Hon Hai Research Institute, Taipei 11492, Taiwan}

\author{Y.-H.~Lin}
\email{yhlin@phys.nthu.edu.tw}
\affiliation{Department of Physics, National Tsing Hua University, Hsinchu 30013, Taiwan}
\affiliation{Center for Quantum Technology, National Tsing Hua University, Hsinchu 30013, Taiwan}

\author{I.-C.~Hoi}
\email{iochoi@cityu.edu.hk}
\affiliation{Department of Physics, City University of Hong Kong, Tat Chee Avenue, Kowloon, Hong Kong SAR, China}
\affiliation{Department of Physics, National Tsing Hua University, Hsinchu 30013, Taiwan}
\date{\today}

\begin{abstract}

A coherent electromagnetic field can be described by its amplitude, frequency, and phase.
All these properties can influence the interaction between the field and an atom.
Here we demonstrate the phase shaping of microwaves that are scattered by a superconducting artificial atom coupled to the end of a semi-infinite 1D transmission line.
In particular, we input a weak exponentially rising pulse with phase modulation to a transmon qubit.
We observe that field-atom interaction can be tuned from nearly full interaction (interaction efficiency, i.e., amount of the field energy interacting with the atom, of \unit[94.5]{\%}) to effectively no interaction (interaction efficiency \unit[3.5]{\%}).

\end{abstract}

\maketitle


\section{Introduction}

Quantum networks, which consist of quantum nodes and quantum channels, have become an important and active research field in recent years~\cite{Kimble2008, Hermans2022}.
To transfer quantum information (e.g., encoded in photons) between the quantum nodes (e.g., atoms), such that it can be processed there, requires interaction between photons and atoms.
In three-dimensional (3D) free space, interaction between propagating photons and atoms is very weak, due to spatial mode mismatch~\cite{Tey2008}.
However, there has been much progress in creating strong interaction between atoms and photons in one-dimensional (1D) space; this field is known as waveguide quantum electrodynamics (QED)~\cite{Roy2017, Gu2017, Sheremet2023}.
In particular, waveguide QED with superconducting artificial atoms~\cite{Gu2017, Krantz2019, Blais2021} and propagating resonant microwave photons has demonstrated such strong interaction in many experiments~\cite{Astafiev2010, Hoi2011, Hoi2012, VanLoo2013, Hoi2013a, Hoi2015, Dmitriev2017, Forn-Diaz2017, Wen2018, Wen2019, Mirhosseini2019, Wen2020, Kannan2020, Vadiraj2020, Joshi2022, Kannan2023}.

In a recent experiment in the setting of waveguide QED, deterministic loading of a resonant microwave pulse onto an artificial atom was achieved~\cite{WeiJuLin2022}.
To further control the interaction between artificial atoms and photons for applications such as quantum networking, quantum sensing~\cite{Degen2017}, transduction of single photons~\cite{Lauk2020}, etc., a switch for tuning the strength of the interaction is necessary.
Currently, a common method for turning on and off the interaction is to use a tunable coupling element~\cite{Kurpiers2018, Axline2018, Campagne-Ibarcq2018, Daiss2021, Kannan2023}.
However, the complex circuit structure of such a coupling element may introduce unwanted modes that cause decoherence for artificial atoms.
We provide an alternative method for tunable coupling in a quantum network.
In this Letter, we use phase shaping~\cite{Specht2009} to continue our previous work~\cite{WeiJuLin2022} and show that the interaction between the field and the atom can be tuned from being fully on with the interaction efficiency up to \unit[94.5]{\%} to effectively being turned off with interaction efficiency down to \unit[3.5]{\%}, where the interaction efficiency indicates how much of the field energy interacts with the qubit.

\begin{figure*}
\includegraphics[width=\linewidth]{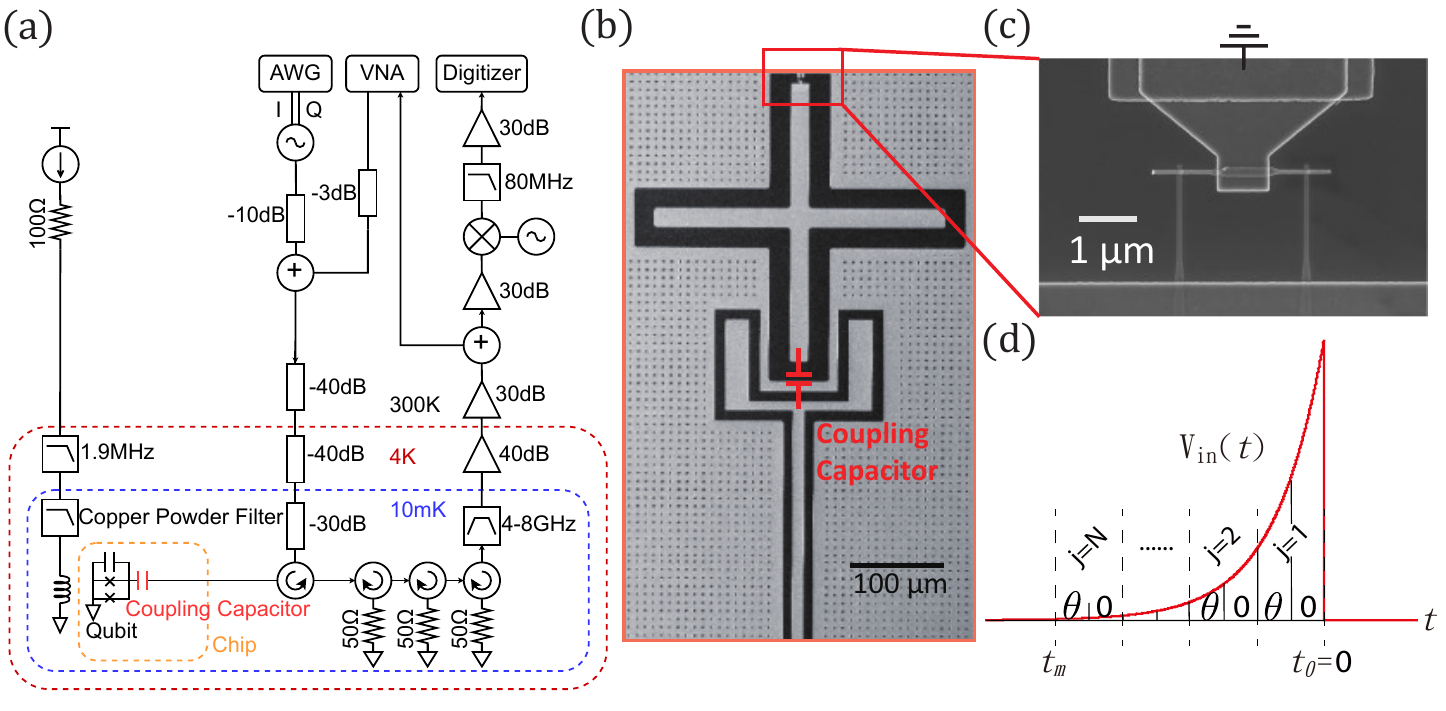}
\caption{
Experimental setup.
(a) Diagram of the full setup. The qubit (a transmon~\cite{Koch2007}) is capacitively coupled to the end of the TL. A vector network analyzer (VNA) for the spectroscopic measurement is connected in parallel to the time-domain measurement system, which consists of a digitizer, an arbitrary waveform generator (AWG), and radio-frequency (RF) sources. The AWG sends in-phase and quadrature (IQ) signals to an IQ modulator (RF source) to generate a phase-shaped pulse. The reflected pulse from the fridge is down-converted by the local oscillator (RF source) and recorded by the digitizer. The recorded data are then sent to a computer for demodulation.
(b) Optical microscope image of the chip layout. The transmon Josephson junctions are located at the top side of the image (red box).
(c) A scanning electron microscope (SEM) image of the transmon Josephson junctions, a superconducting quantum interference device (SQUID), which allows tuning the transmon resonance frequency by an external magnetic field. The upper patch belongs to the ground plane while the lower patch belongs to the charge island.
(d) The exponentially rising waveform, $V_{\rm in}(t)$, with a \unit[50]{\%} duty cycle and $0-\theta$ phase shaping, as defined in Eqs.~(\ref{Eq:exprPulse}) and (\ref{Eq:phaseMod}), is generated by the AWG and the IQ modulator.
\label{fig:setup}}
\end{figure*}

In particular, we send a weak exponentially rising coherent pulse with phase modulation towards a superconducting artificial atom in a semi-infinite 1D transmission line (TL), as depicted in \figref{fig:setup}.
We achieve coherent control of the interaction by manipulating the phase of the coherent input state.
By interleaving segments with phases 0 and $\theta$ in the exponentially rising pulse, as illustrated in \figpanel{fig:setup}{d}, the rotational axis of the qubit state changes during the excitation process.
For $\theta = \pi$ with a large number of segments $N$, almost no interaction will occur.
By varying $\theta$, we are thus able to tune the interaction between the photon and the qubit.

\section{Measurement}

We first characterize our sample using reflective spectroscopy with a vector network analyzer (VNA) and extract necessary parameters (e.g., qubit resonance frequency $\omega_{10}$, radiative relaxation rate $\Gamma$, and decoherence rate $\gamma$) to be used in the time-domain measurements and simulations.
The extraction method and the extracted parameters are presented in \appref{app:calibrations}.

For time-domain measurements we use an arbitrary waveform generator (AWG) and a radio-frequency (RF) source with in-phase and quadrature (IQ) modulation capability [see \figpanel{fig:setup}{a}] to generate the phase-modulated exponentially rising pulse with the envelope voltage
\begin{equation}
	V_{\text{in}}(t) = V \Theta(t_0 - t) e^{(t-t_0) / \tau} e^{i \Pi(t)},
	\label{Eq:exprPulse}
\end{equation}
where $V$ is the peak magnitude of the input voltage at the chip, $\Theta(t)$ is the Heaviside step function, $t_0 = 0$ is the time when the pulse reaches its maximum and is turned off, $\tau$ is the characteristic time of the exponentially rising waveform, and $\Pi(t)$ is a \unit[50]{\%} duty-cycle pulse train [here we define the duty cycle as the ratio of the $\theta$-interval time span and the pulse period in $\Pi(t)$]. The pulse train is responsible for the phase shaping, which is given by
\begin{equation}
	\Pi(t) =
	\begin{cases}
		\theta & t_0 - j \Delta t \leq t < t_0 - (j - \frac{1}{2}) \Delta t
		\\0 & \text{elsewhere}
	\end{cases},
	\label{Eq:phaseMod}
\end{equation}
where $j = 1, 2, \ldots, N$ represents the $j$th interval as shown in \figpanel{fig:setup}{d}, $\theta \in [0, 2\pi]$ is the modulated phase, and $\Delta t = (t_0 - t_m) / N$ is the switching period of the modulated phase determined by the number of intervals $N$,  the modulation start time $t_m$, and the end time $t_0$.
In order to observe sufficient case variations for $N \in [0, 50]$, $t_m$ is set to $\unit[-2.5]{\mu s}$. 

In our previous work~\cite{WeiJuLin2022}, we demonstrated that perfect atom-photon interaction for an exponentially rising pulse occurs for a weak input field [$\Omega(t) \ll \gamma$, where $\Omega(t)$ is the Rabi frequency with maximum magnitude $\Omega$] when the characteristic time $\tau$ of the pulse equals the decoherence time $T_2 = 1 / \gamma$, such that the input waveform has the same shape as the time-reversed qubit emission.
Throughout this manuscript we set $\Omega / 2 \pi \approx \unit[0.154]{MHz}$, which is about 10 times less than $\gamma$, and we also set $\tau = T_2$.

\begin{figure*}
\includegraphics[width=\linewidth]{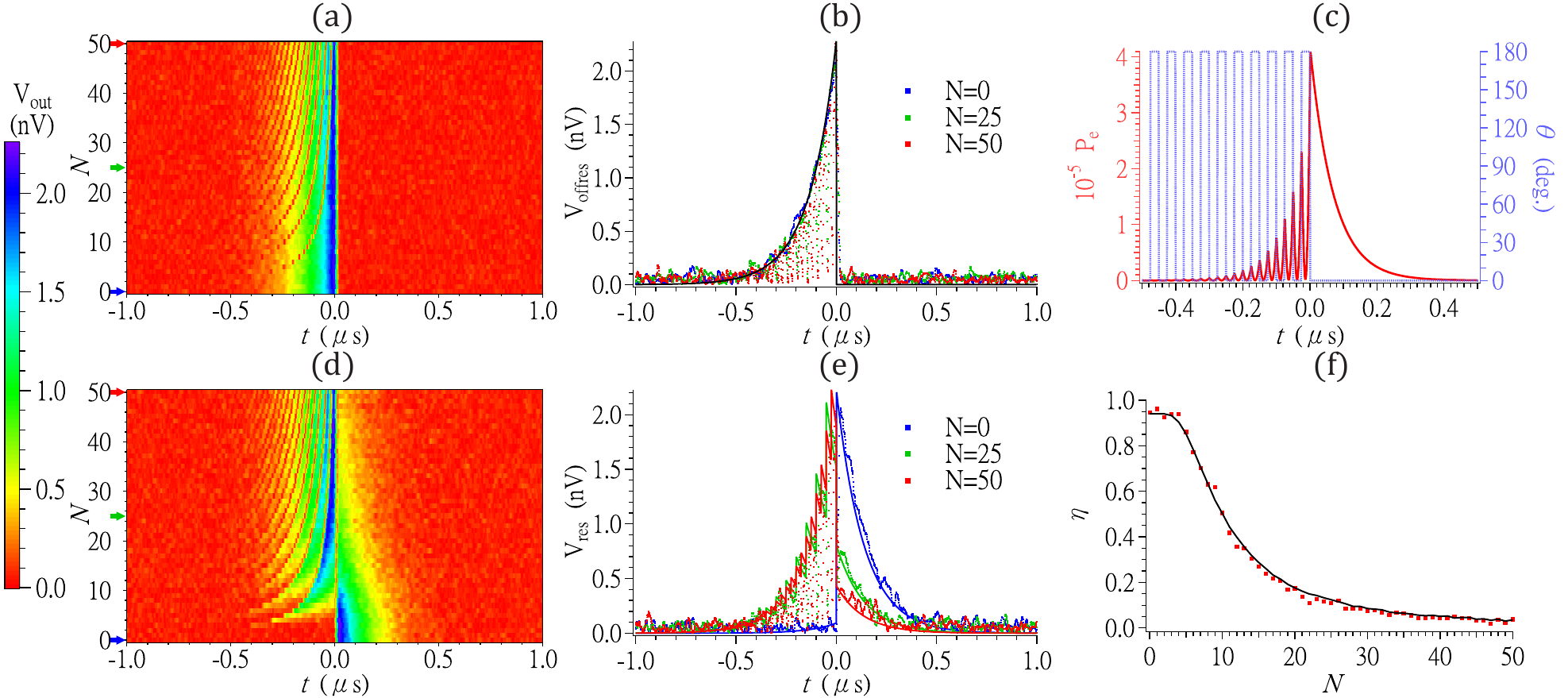}
\caption{Scattering of an exponentially rising pulse with $N \in [0, 50]$ and fixing $\theta = \pi$.
The measured data is presented as dots and theoretical simulations as solid curves.
The simulations are done by numerically solving Eqs.~(\ref{Eq:io})--(\ref{Eq:obe_z}) with parameters extracted from the spectroscopy, as indicated in \tabref{tab:Parameters} in \appref{app:calibrations}; no free fitting parameter is assigned.
(a) Reflected pulse envelope voltage with the qubit far detuned.
(b) Cross sections of (a) for the cases $N=0$ (blue), 25 (green), and 50 (red).
(c) Simulated occupation probability $P_e$ of the qubit's first excited state and corresponding modulated phase ($N=50$, $\theta = \pi$) as functions of time.
(d) Reflected pulse envelope voltage with the qubit on resonance with the probe.
(e) Cross sections of (d) for the cases $N=0$ (blue), 25 (green), and 50 (red).
(f) $\eta$ as a function of $N$.
The maximum $\eta$ at $N = 0$ is $\unit[94]{\%} \pm \unit[1.0]{\%}$, which matches well with the value predicted by the analytic formula (\unit[93.5]{\%}).
For the $N=50$ case the measured and simulated $\eta$ are $\unit[3.8]{\%} \pm \unit[1.0]{\%}$ and \unit[3.6]{\%}, respectively.
In (b,e), the dips appearing in the exponentially rising pulse are caused by the finite bandwidth of the digitizer, which limits the demodulation time to \unit[20]{ns} (corresponding to \unit[50]{MHz} demodulation frequency).
The non-zero demodulation time smooths out data points and results in dips.
These dips can also be found in panels (a) and (d) as near-zero-voltage strips between $0$ and $\theta$ intervals before $t_0$.
Full simulations of panels (a) and (d) are shown in \figref{fig:sim_2D_plots} in \appref{app:SimulatedReflection}.
}
\label{fig:swp_N}
\end{figure*}

\begin{figure*}
\includegraphics[width=\linewidth]{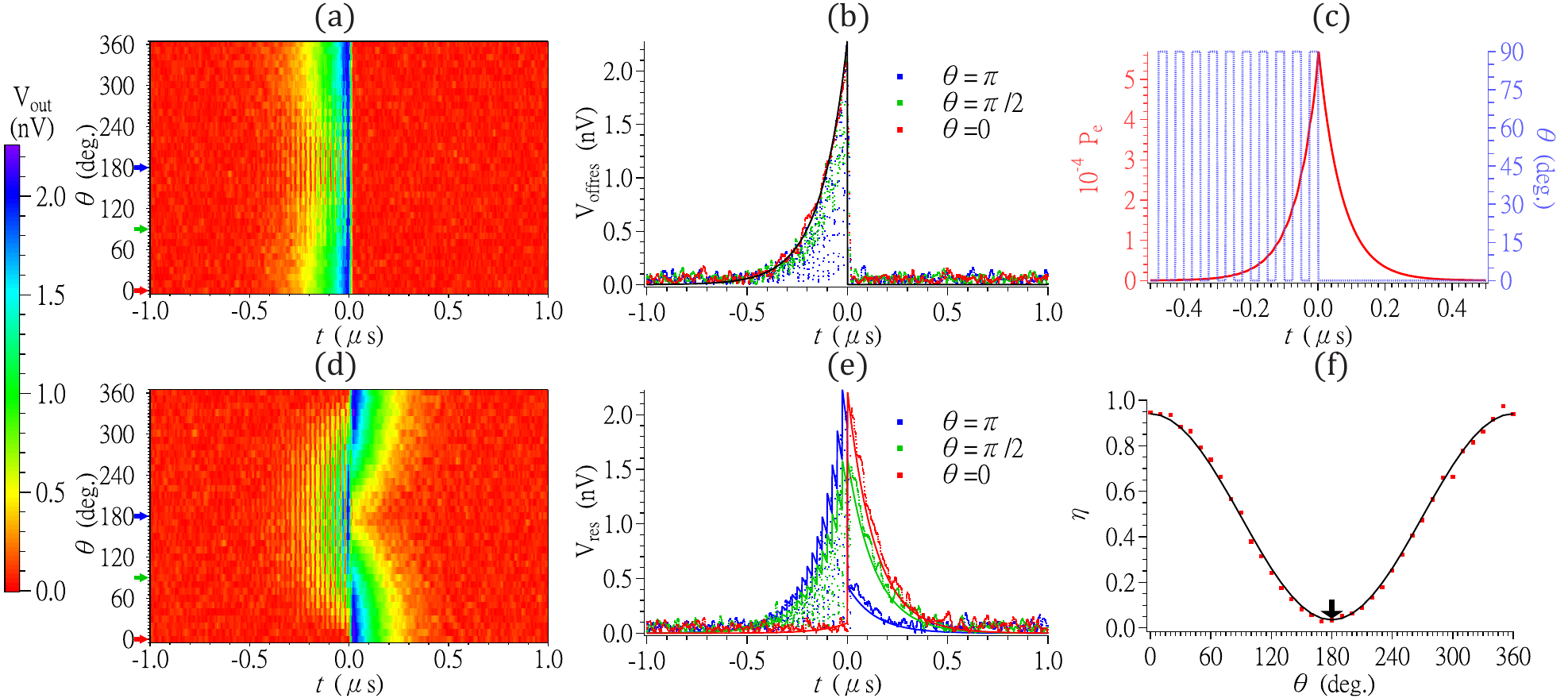}
\caption{Scattering of an exponentially rising pulse, fixing $N = 50$ and sweeping $\theta$ from 0 to $2\pi$. The measured data is presented as dots and theoretical simulations as solid curves.
(a) Reflected pulse envelope voltage with the qubit far detuned.
(b) Cross sections of (a) for the cases $\theta = 0$ (red), $\pi/2$ (green), and $\pi$ (blue).
(c) First-excited-state occupation probability for the qubit and corresponding modulated phase ($N = 50$, $\theta = \pi/2$) as functions of time. Note that $P_e$ is nearly symmetric with respect to $t_0$.
(d) Reflected pulse envelope voltage with the qubit on resonance with the probe.
(e) Cross sections of (d) for the cases $\theta = 0$ (red), $\pi/2$ (green), and $\pi$ (blue).
(f) $\eta$ as a function of $\theta$.
At $\theta = \pi/2$, $\eta$ is at the middle of its range ($\unit[50.6]{\%} \pm \unit[1.0]{\%}$ measured, \unit[48.8]{\%} simulated).
For $\theta = 0$ the interaction suppression is off and $\eta$ is at its maximum ($\unit[94.5]{\%} \pm \unit[1.0]{\%}$ measured, \unit[94]{\%} simulated).
Conversely, $\eta$ reaches its minimum when $\theta = \pi$ ($\unit[3.5]{\%} \pm \unit[1.0]{\%}$ measured, \unit[3.6]{\%} simulated) indicated by the black arrow.
Full simulations of panels (a) and (d) are shown in \figref{fig:sim_2D_plots} in \appref{app:SimulatedReflection}.
}
\label{fig:swp_theta}
\end{figure*}

The output voltage $V_{\text{out}}(t)$ is given by input-output theory~\cite{Lalumiere2013}:
\begin{equation}
	V_{\text{out}}(t) = V_{\text{in}}(t) + \frac{2 \Gamma}{k} \sqrt{2 Z_0} \expec{\sm (t)},
	\label{Eq:io}
\end{equation}
where $Z_0 = \unit[50]{\Omega}$ is the characteristic impedance of the TL and $k$ is the proportionality constant relating $\Omega(t)$ and the square root of the input power $P_{\text{in}}(t) = \abssq{V_{\text{in}}(t)} / 2 Z_0$; $\Omega(t) = k \sqrt{P_{\text{in}}(t)}$.
In \eqref{Eq:io}, the coherent output field receives contributions from terms representing the input field and the atomic emission.
The atomic term consists of the expectation value of the Pauli lowering operator $\sm$, whose time evolution is described by the optical Bloch equations~\cite{WeiJuLin2022}
\begin{align}
	\partial_t \expec{\sp} &= (-i \delta - \gamma) \expec{\sp} + \Omega^*(t) \expec{\sz} / 2,
	\label{Eq:obe_p}
	\\
	\partial_t \expec{\sm} &= (i\delta - \gamma) \expec{\sm} + \Omega(t) \expec{\sz} / 2,
	\label{Eq:obe_m}
	\\
	\partial_t \expec{\sz} &= -\Gamma (1 + \expec{\sz}) - \Omega(t) \expec{\sp} - \Omega^*(t) \expec{\sm},
	\label{Eq:obe_z}
\end{align}
where $\sp$ and $\sz$ are the Pauli raising and Z operators, respectively, and $\delta$ represents the detuning between the input signal frequency $\omega_p$ and the qubit transition frequency, i.e., $\delta = \omega_p - \omega_{10}$. 
We numerically solve Eqs.~(\ref{Eq:io})--(\ref{Eq:obe_z}) based on the extracted qubit parameters from the frequency-domain measurement detailed in \appref{app:calibrations} (see \tabref{tab:Parameters} there).

To quantify the effectiveness of the interaction, we define the input energy $E_{\rm offres}$ (measured when the qubit is far detuned), the output energy $E_{\rm res}$ (measured when the probe is on resonance with the qubit), and the coherent interaction efficiency $\eta = E_{\rm res} / E_{\rm offres}$~\cite{WeiJuLin2022} with  
\begin{flalign}
	E_{\rm offres} &= \frac{1}{2Z_0} \int_{t_i}^{t_0} [\abssq{V_{\rm offres}(t)} - \abssq{V_N}] dt,
	\label{Eq:energy_off}
	\\
	E_{\rm res} &= \frac{1}{2Z_0} \int_{t_0}^{t_f} [\abssq{V_{\rm res}(t)} - \abssq{V_N}] dt,
	\label{Eq:energy_on}
\end{flalign}
where $V_{\rm res}$ ($V_{\rm offres}$) represents the measured $V_{\rm out}$ at the chip level when the qubit is tuned on (far off) resonance with the probe tone, $V_N$ denotes the average noise level over the time interval $t \in [\unit[1]{\mu s}, \unit[5]{\mu s}]$ after turning off the input pulse in the off-resonant case ($V_{\rm offres}$), $t_i$ is the pulse start time, and $t_f$ is the measurement stop time. A \unit[100]{\%} interaction efficiency means that the energy of the coherent output field is equal to the energy of the input field.

There are two driving regimes for the exponentially rising waveform: weak driving and strong driving.
The weak driving regime, where $\Omega(t) \ll \gamma$, is the one investigated throughout this work.
In this regime, all the input field is elastically scattered, leading to a nearly \unit[100]{\%} efficiency; the qubit is mostly in the ground state. Using Eqs.~(\ref{Eq:exprPulse}) and (\ref{Eq:energy_off}) and the selected $V \approx \unit[2]{nV}$, we see that the exponentially rising waveform contains an average photon number $E_{\rm offres}/(\hbar \omega_{10}) \approx 0.0011$.
On the other hand, in the strong driving regime $\Omega(t) \gg \gamma$, the microwaves are both elastically and inelastically scattered, leading to missing energy in other frequencies and therefore a lower $\eta$.

From the analytic formula (derived in the limit of a weak probe, $\Omega(t) \ll \gamma$)~\cite{WeiJuLin2022}
\begin{equation}
	\eta = \frac{\Gamma^2 / \tau}{(\frac{\Gamma}{2} + \Gamma_{\phi,l})(\frac{\Gamma}{2} + \Gamma_{\phi,l} + 1/ \tau)^2},
	\label{Eq:analytic_eff}
\end{equation}
we obtain and estimate a maximum interaction efficiency of $\eta_{\rm max} \approx \unit[93.5]{\%}$ at $\tau = T_2$.
From \eqref{Eq:analytic_eff}, we see that $\eta_{\rm max}$ is limited by $\Gamma_{\phi,l} = \Gamma_{\phi} + \Gamma_{\rm nr}/2$, where $\Gamma_{\phi}$ is the qubit's pure dephasing rate and $\Gamma_{\rm nr}$ its non-radiative relaxation rate~\cite{hoi2013microwave}.
Although our measurement method cannot separate $\Gamma_{\rm nr}$ from $\Gamma_{\phi}$, we can use the extracted $\Gamma_{\phi,l}$ in \tabref{tab:Parameters} in \appref{app:calibrations} to estimate the maximum $\Gamma_{\rm nr}$ to be $2\Gamma_{\phi,l} \approx \unit[77]{kHz}$, which is $\unit[3.4]{\%}$ of $\Gamma$.

Losses are defined as energy that is not reflected coherently; this includes incoherent scattering and non-radiative relaxations.
Due to energy conservation, the sum of these power losses is given by $P_{\rm loss} = P_{\rm in}(1-|r|^2)$~\cite{Lu2021}, where the reflection coefficient $r$ is given in \eqref{Eq:refl_coeff} in \appref{app:calibrations}.
In the steady state of constant wave excitation at $\Omega / 2 \pi \approx \unit[0.154]{MHz}$, $P_{\rm loss}$ is \unit[15.8]{\%} of $P_{\rm in}$.


\section{Results from sweeping the number of intervals}
\label{sec:sec_swp_interval}

As shown in \figref{fig:swp_N}, we sweep $N$ from 0 to 50 (fixing $\theta = \pi$) and observe that increasing $N$ leads to incresing suppression of the interaction efficiency.
We use Eqs.~(\ref{Eq:energy_off})--(\ref{Eq:energy_on}) and the data in \figpanel{fig:swp_N}{a,d} to calculate the result for the efficiency as a function of $N$ shown in \figpanel{fig:swp_N}{f}~\cite{WeiJuLin2022}.

The principle behind the interaction suppression via phase shaping can be understood from \figpanel{fig:swp_N}{c}.
The accumulated occupation probability $P_e$ for the excited state of the qubit during the $\theta = 0$ period is cancelled by the adjacent $\theta = \pi$ period, which inverts the rotational axis of the Bloch vector.
This makes the Bloch vector swing back and forth around the ground state ($\expec{\sz} = -1$) during the pulse.
With the increase of $N$, the time for the qubit excitation is shortened and thus the interaction is suppressed as shown in \figpanel{fig:swp_N}{f}.
However, to reach total interaction suppression one may need a very large $N$.
It can be seen in \figpanel{fig:swp_N}{f} that the slope of $\eta$ gradually decreases as $N$ increases.
This makes complete interaction suppression hard to be achieved with a \unit[50]{\%} duty cycle, since the maximum possible $N$ is limited by the AWG sampling rate and the IQ modulator's input bandwidth.
The use of linear phase modulation to accommodate equipment bandwidth is discussed in \appref{app:SawtoothModulation}.

Theoretically, it is however possible to achieve zero $\eta$ for our experimental parameters by tuning the duty cycle to \unit[59.1]{\%}, as calculated in \appref{app:DutyCycleTuning}. This optimum comes from considering both that the \unit[50]{\%} duty cycle introduces a difference in the total areas of the 0 and $\pi$ intervals, and the effect of decoherence.


\section{Results from sweeping the modulation phase}

In \figref{fig:swp_theta}, we sweep $\theta$ from 0 to $2 \pi$ to tune the interaction suppression while fixing $N = 50$ and the duty cycle to \unit[50]{\%}.
This sweep effectively rotates the direction of the rotational axis on the equatorial plane of the Bloch sphere by $\theta$ and allows us to steer the direction of the Bloch-vector evolution during input.

Zero (maximum) interaction suppression is achieved when $\theta = 0$ $(\pi)$.
For $0 < \theta < \pi/2$, the $\theta$ interval provides partial boosting of $P_e$ and results in an $\eta$ between \unit[50]{\%} and \unit[100]{\%}.
In contrast, $\pi/2 < \theta < \pi$ partially suppresses the interaction such that $\eta$ is between \unit[0]{\%} and \unit[50]{\%}.
As a balance point between these two intervals, the case $\theta = \pi/2$ has nearly (due to finite $N$) \unit[50]{\%} interaction efficiency, as shown in \figpanel{fig:swp_theta}{f}. 
The corresponding $P_e$ as a function of time is depicted in \figpanel{fig:swp_theta}{c}.
The remaining $\pi < \theta < 2\pi$ cases are the mirror images of $0 < \theta < \pi$.
From the results in \figpanel{fig:swp_theta}{f}, we see that we can easily tune the atom-field interaction to any desired $\eta$ value between the maximum and minimum on demand by setting $\theta$.


\section{Conclusion}

We demonstrated phase shaping of microwaves being scattered by a superconducting artificial atom in a semi-infinite 1D transmission line in time domain.
In particular, we sent in a weak exponentially rising pulse with phase modulation towards the atom and observed that the atom-field interaction can be tuned from nearly full interaction to effectively no interaction, as measured by the amount of energy transferred from the field to the atom (the interaction efficiency).
The maximum interaction efficiency can be increased by improving fabrication process (lowering pure dephasing and non-radiative relaxation rates).
To improve interaction cancellation, there are two routes to take, which also can be used in combination: tuning the duty cycle of the pulse and tuning the number of phase-switching intervals $N$.
Our results may enable promising applications, through tunable interaction, in quantum networks based on waveguide quantum electrodynamics.


\section{Acknowledgements}

I.-C.H.~and J.C.C.~thank I.A.~Yu for fruitful discussions.
I.-C.H.~acknowledges financial support from City University of Hong Kong through the start-up project 9610569 and from the Research Grants Council of Hong Kong (Grant No.~11312322).
A.F.K.~acknowledges support from the Swedish Research Council (grant number 2019-03696), from the Swedish Foundation for Strategic Research, and from the Knut and Alice Wallenberg Foundation through the Wallenberg Centre for Quantum Technology (WACQT).
K.-T.L. and G.-D.L.~acknowledge support from NSTC of Taiwan under Projects No. 111-2112-M-002-037 and 111-2811-M-002 -087.
P.Y.W.~acknowledges support from MOST, Taiwan, under grant No. 110-2112-M-194-006-MY3.


\appendix

\section{Calibrations}
\label{app:calibrations}

As depicted in \figpanel{fig:setup}{a} in the main text, we can generalize both frequency- and time-domain setups into a simplified scheme: an RF source launches a pulse with power $P_{\text{src}} = V_{\text{src}}^2 / 2 Z_0$ (the subscript `src' refers to the source) towards the qubit via an effective attenuator with power attenuation $A$.
The qubit has a reflection coefficient $r(\delta, \Omega) = V_{\rm out}(\infty) / V_{\rm in}(\infty)$, which can be found from the stationary solution of Eqs.~(\ref{Eq:io})--(\ref{Eq:obe_z}):
\begin{equation}
	r(\delta, \Omega) = 1 - \frac{\Gamma}{\gamma} \frac{1 - i \frac{\delta}{\gamma}}{1 + \mleft(\frac{\delta}{\gamma} \mright)^2 + \frac{\Omega ^2}{\gamma \Gamma}},
	\label{Eq:refl_coeff}
\end{equation}
where $\Omega$ is the continuous-wave (CW) Rabi frequency and $\gamma = \Gamma / 2 + \Gamma_{\phi,l}$.

The reflected signal passes through the amplifier chain (with effective power gain $G$) and finally reaches the receiver with voltage
\begin{equation}
	V_{\text{rec}} = \sqrt{G} r(\delta, \Omega) \sqrt{A} V_{\text{src}}.
	\label{Eq:sensingChain}
\end{equation}
Here we assume no multiple reflections occur in our transmission so that \eqref{Eq:sensingChain} applies.
The measured reflection coefficient is defined as
\begin{equation}
	r_{\text{all}} = \frac{V_{\rm rec}}{V_{\rm src}}.
	\label{Eq:refl_all}
\end{equation}

\begin{figure*}
\includegraphics[width=\linewidth]{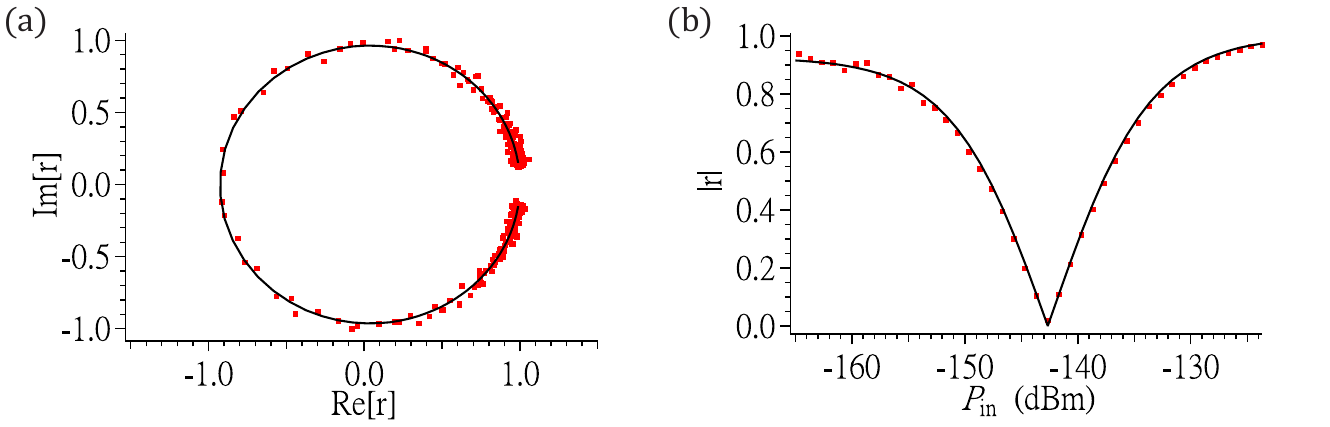}
\caption{Spectroscopy results.
(a) Reflection-coefficient IQ plot at $\unit[-163]{dBm}$ ($\Omega \ll \gamma$).
(b) On resonance ($\delta = 0$) power-dependent reflection magnitude.
In both plots, the red dots are the measured data points and the solid curves (black) are the theory fitting according to \eqref{Eq:refl_coeff}.
}
\label{fig:qb_Chara}
\end{figure*}

We first sweep the probe frequency near the qubit resonance frequency with sufficiently low power ($\Omega \ll \gamma$) such that $r(\delta, \Omega \approx 0)$ is nearly a Lorentzian.
It is straightforward to extract $r(\delta, \Omega)$ by dividing \eqref{Eq:sensingChain} with the far-detuned case (i.e., background) where $r(\delta \rightarrow \infty, \Omega) \approx 1$ and thus $r_{\text{all,bg}} = \sqrt{GA}$.
The extracted $r(\delta, \Omega \approx 0)$ is shown in \figpanel{fig:qb_Chara}{a}.
For the Lorentzian function we use the circle-fit method~\cite{probst2015efficient, Lu2021} to extract $\Gamma$ and $\gamma$ and $\omega_{10}$, which are summarized in \tabref{tab:Parameters}.

\begin{table*}
	\centering
	\begin{tabular}{| c | c | c | c | c | c | c | c | c |}
		\hline
		$\omega_{10} / 2 \pi$&
		$\Gamma / 2 \pi$&
		$\gamma / 2 \pi$&
		$\Gamma_{\phi,l}/2\pi$&
		$A_{\rm spec}$&
		$G_{\rm spec}$&
		$A_{\rm time}$&
		$G_{\rm time}$\\
		\hline
		[MHz]&
		[MHz]&
		[MHz]&
		[MHz]&
		[dB]&
		[dB]&
		[dB]&
		[dB]\\
		\hline
		$4766 \pm 0.010$ &
		$2.271 \pm 0.013$ &
		$1.174 \pm 0.010$ &
		$0.038 \pm 0.012$ &
		$-133.66 \pm 0.03$ &
		$60.87 \pm 0.03$ &
		$-154.84 \pm 0.03$ &
		$104.51 \pm 0.04$ \\
		\hline
	\end{tabular}
	\caption{Extracted and derived qubit and setup parameters at $\omega_{10}$. $\Gamma_{\phi,l}$ is calculated using $\Gamma_{\phi,l} = \gamma - \Gamma / 2$. The subscripts for $A$ and $G$ are used to distinguish between spectroscopy and time-domain systems.}
	\label{tab:Parameters}
\end{table*}

The next step is to calibrate the constants $k$, $A$, and $G$.
By tuning the probe on resonance with the qubit ($\delta = 0$) and sweeping $\Omega$, we obtain the power-dependent reflection coefficient $r(\delta=0, \Omega)$ [\figpanel{fig:qb_Chara}{b}] after the background is removed.
We define another proportionality constant $k_{\text{src}}$ for $P_{\text{src}}$:
\begin{equation}
	\Omega = k_{\text{src}} \sqrt{P_{\text{src}}}.
	\label{Eq:k_src_def}
\end{equation}
This constant can be obtained by fitting $r(\delta=0, \Omega)$ with $\sqrt{P_{\text{src}}}$ via \eqref{Eq:refl_coeff}.
Comparing to the definition of $k$ used in the main text,
\begin{equation}
	\Omega (t) = k \sqrt{P_{\text{in}}(t)},
	\label{Eq:k_def}
\end{equation}
and due to the fact that $P_{\text{in}}(t) = A P_{\text{src}}$, the two constants are related by
\begin{equation}
	k_{\text{src}} = \sqrt{A} k.
	\label{Eq:k_src_k_relation}
\end{equation}
To extract $A$, an algebraic identity between $\Gamma$, $P_{src}$, and $\Omega$, proven in the Supplementary Material of Ref.~\cite{Hoi2015}, is used:
\begin{equation}
	\Omega = \sqrt{\frac{8 \pi \Gamma}{\hbar \omega_r}} \sqrt{AP_{\rm src}},
	\label{Eq:relax_idd}
\end{equation}
where $\hbar$ is the reduced Planck's constant and $\omega_r$ is the qubit's resonance frequency. From Eqs.~(\ref{Eq:k_src_def}) and (\ref{Eq:relax_idd}) we arrive at
\begin{equation}
	A = k_{\text{src}}^2 \hbar \omega_r / 8 \pi \Gamma.
	\label{Eq:Attenuation}
\end{equation}
Then, using \eqref{Eq:k_src_k_relation}, we obtain the expression
\begin{equation}
	k = \sqrt{8 \pi \Gamma / \hbar \omega_r}.
	\label{Eq:k_expr}
\end{equation}
At last the gain $G$ is obtained immediately:
\begin{equation}
	G = \abssq{r_{\rm all,bg}} / A.
	\label{Eq:Gain}
\end{equation}
Here only the magnitude of $r_{\rm all,bg}$ is used because the round-trip phase due to TL and microwave components can be removed by dividing reflection coefficients in the first step.
With extracted $\Gamma$, $\gamma$, $k$, $A$, and $G$, we can simulate the time-domain results using Eqs.~(\ref{Eq:io})--(\ref{Eq:obe_z}) without the use of any free parameter.


\section{Simulated zero interaction efficiency by duty cycle tuning}
\label{app:DutyCycleTuning}

\begin{figure}
\includegraphics[width=\linewidth]{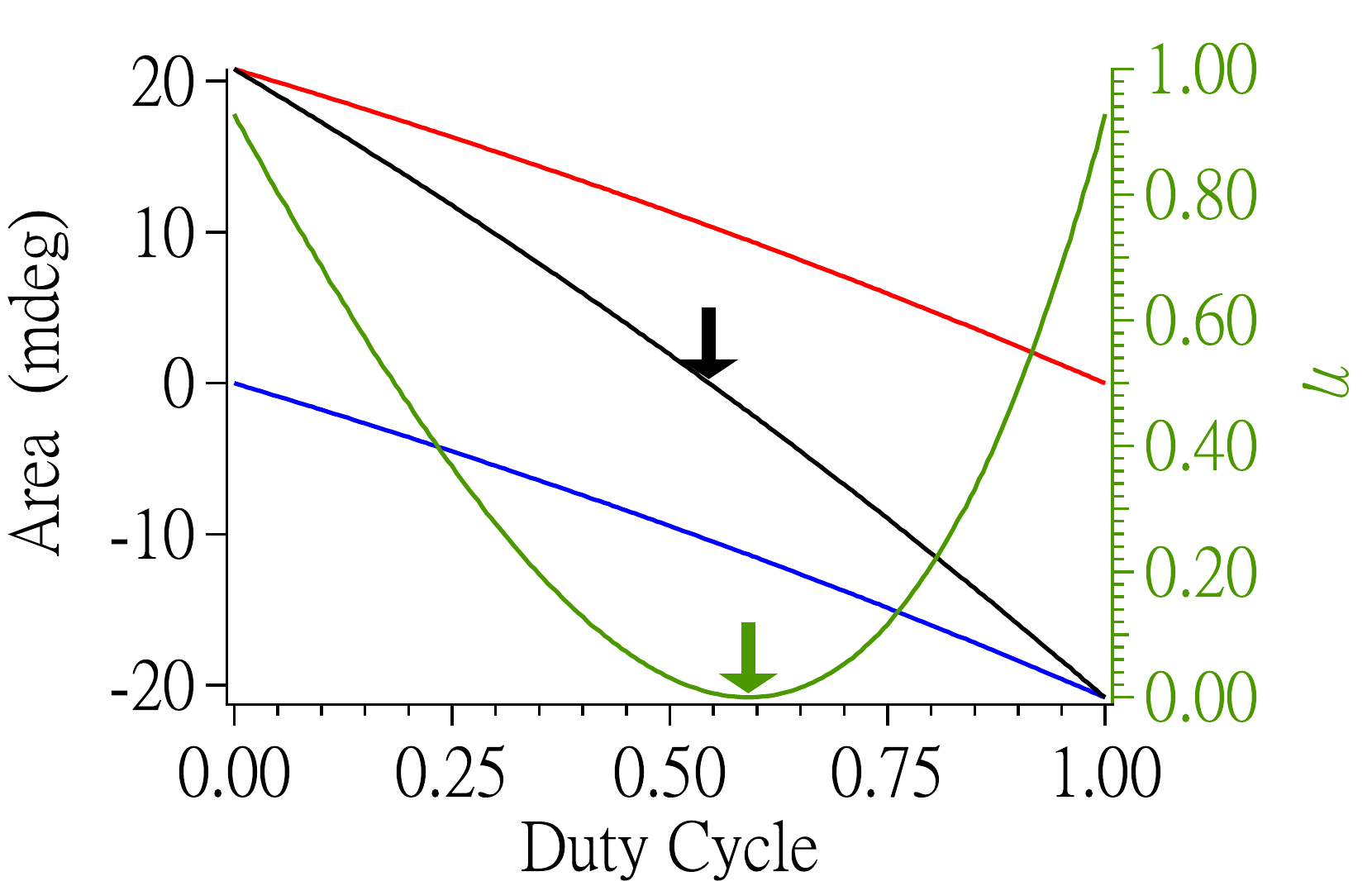}
\caption{Simulated results for changing duty cycle when $N = 50$ and $\theta = \pi$. The red(blue) line is the total angular area swept by the $0$ ($\pi$)-degree part of the time-varying Rabi frequency. The green curve is $\eta$ as a function of duty cycle. The sum of the two areas (black line) is $0$ when the duty cycle reaches \unit[54.5]{\%} (black arrow). However, due to the presence of relaxation and decoherence, the actual zero efficiency occurs at \unit[59.1]{\%} (green arrow).}
\label{fig:sim_area_cancel}
\end{figure}

To further suppress $\eta$ beyond what was achieved in \figpanel{fig:swp_N}{f} in the main text, we can optimize the waveform by changing the duty cycle of $\Pi(t)$ and cancel the Bloch-vector rotation induced by the non-uniform Rabi frequency.
Our simulation (\figref{fig:sim_area_cancel}) shows that the optimal duty cycle is \unit[59.1]{\%}.

There are two reasons for the optimal duty cycle not being \unit[50]{\%}.
First, the difference in total areas between $0$ and $\pi$ intervals [see \figpanel{fig:setup}{d}] leads to an additional rotation angle, which corresponds to a residual excited-state population.
For our exponentially rising pulse, a duty cycle of \unit[54.5]{\%} would cancel all $0$ and $\pi$ areas perfectly.
However, secondly, the qubit decoherence appears as a force dragging the qubit towards the ground state, creating additional overshoots for the rotating Bloch vector and pushing the optimum point to \unit[59.1]{\%}.


\section{Simulated reflected signals}
\label{app:SimulatedReflection}

\begin{figure*}
\includegraphics[width=\linewidth]{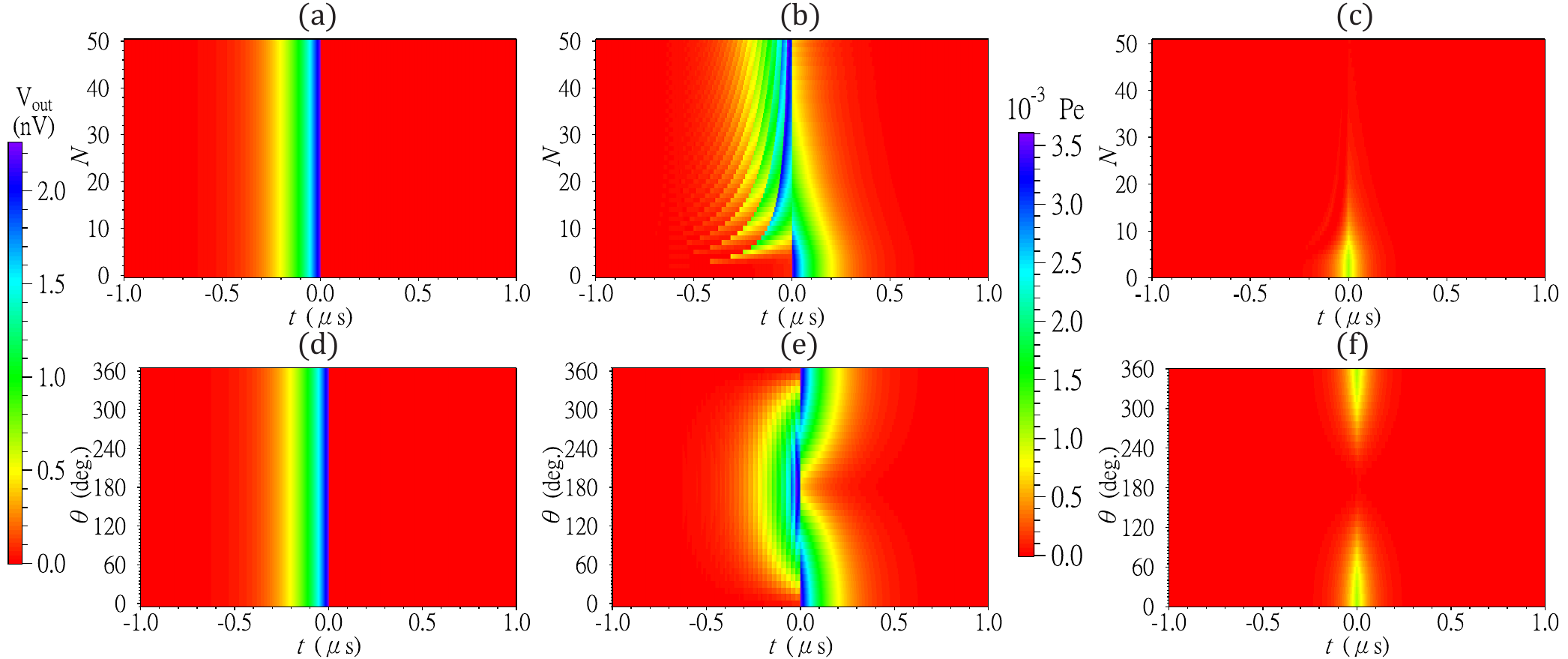}
\caption{Simulated reflected voltages $V_{\rm out} (t)$ and qubit population $P_e (t)$.
The corresponding experimental results are shown in \figref{fig:swp_N} and \figref{fig:swp_theta} in the main text.
(a) Varying $N$ and fixing $\theta = \pi$ as the qubit is far detuned.
(b,c) Varying $N$ and fixing $\theta = \pi$ as the probe tone is on resonance with the qubit.
(d) Varying $\theta$ and fixing $N=50$ as the qubit is far detuned.
(e,f) Varying $\theta$ and fixing $N=50$ as the probe tone is on resonance with the qubit.
}
\label{fig:sim_2D_plots}
\end{figure*}

In \figref{fig:sim_2D_plots}, we show simulation results corresponding to \figpanel{fig:swp_N}{a,d} and \figpanel{fig:swp_theta}{a,d} in the main text. The simulations are done by numerically solving Eqs.~(\ref{Eq:io})--(\ref{Eq:obe_z}) with parameters extracted from the spectroscopy (see \appref{app:calibrations}); no free fitting parameter is assigned. We also show simulation results for the qubit population.

\section{Linear phase modulation}
\label{app:SawtoothModulation}

\begin{figure*}
\includegraphics[width=\linewidth]{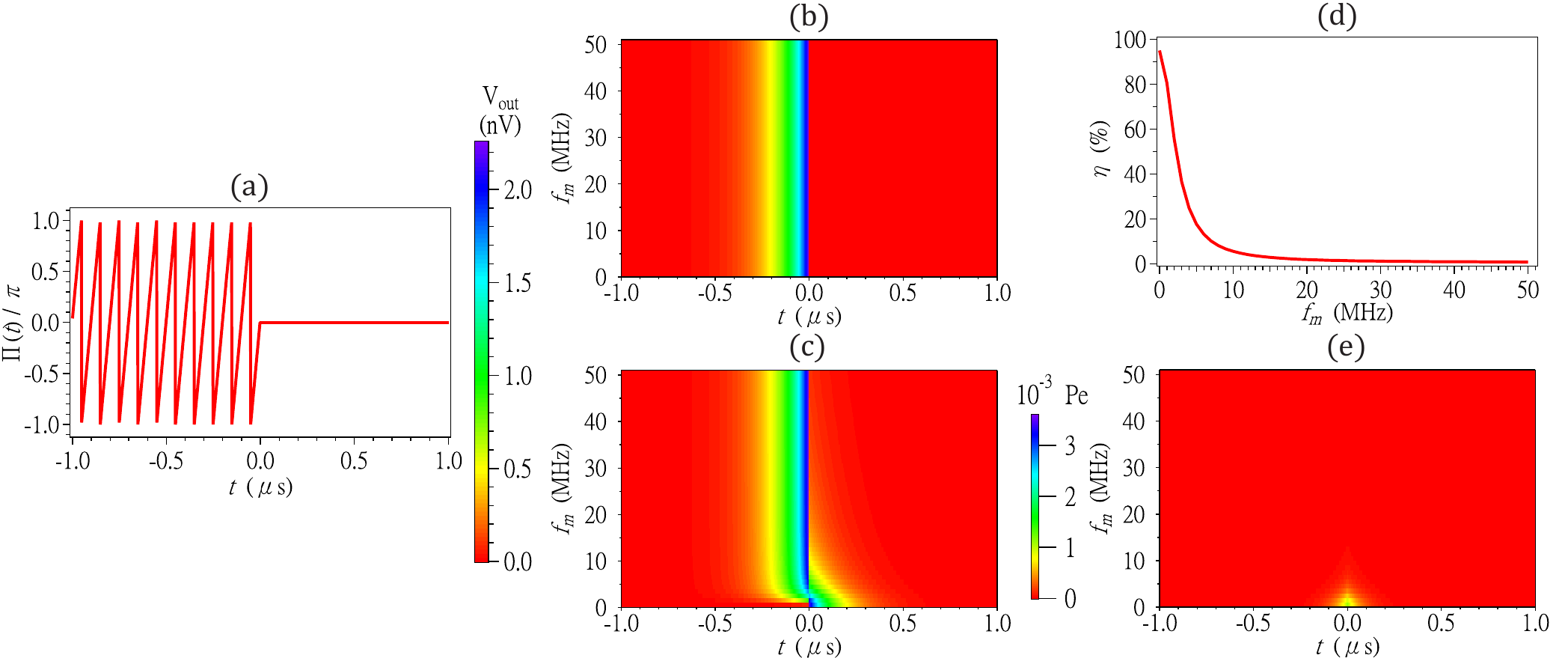}
\caption{Simulated reflected voltages $V_{\rm out} (t)$ and qubit population $P_e$ with sawtooth phase modulation.
(a) Modulation phase as a function of time.
(b) Reflected voltage as a function of modulation frequency $f_m$ and time $t$ when the probe tone is on resonance with the qubit.
(c) Reflected voltage as a function of $f_m$ and time when the qubit is far detuned.
(d) $\eta$ as a function of $f_m$.
(e) Population $P_e$ as a function of time and modulation frequency.
}
\label{fig:Sawtooth_sim}
\end{figure*}

There is a larger family of pulses that can exhibit the kind of cancellation that we use in this work. In general, the phase modulation function $\Pi(t)$ multiplying the exponentially rising pulse should have the following two properties.
First, $\Pi(t)$ must encompass both positive and negative amplitude cycles to effectively counterbalance each other.
Second, the period of $\Pi(t)$ should be short enough to cancel similar pulse amplitudes adjacent in time.
If the period is limited by the instrument bandwidth, an optimization on duty cycle, as discussed in \appref{app:DutyCycleTuning}, is also an option.

The periodicity mentioned in the second property leads us to expand $\Pi(t)$ in a Fourier series.
Taking the square wave in \eqref{Eq:phaseMod} ($N = 50$ and $\theta = \pi$) as an example, the harmonics in the series occur at integer multiples of a frequency of \unit[20]{MHz}, which is the repetition rate of the square wave.
These harmonics generate sideband modulations that detune the carrier signal beyond the qubit linewidth, effectively disabling absorption.
In this sense, a simpler alternative to the square wave is to use a pair of sine and cosine waves on the IQ ports of an IQ modulator.
In terms of phase, this is effectively a linear phase modulation (or sawtooth modulation, if $\Pi(t)$ is limited to being within $[-\pi,\pi]$), expressed as
\begin{equation}
	\Pi(t) = f_mt
	\label{Eq:phaseMod_linear}
\end{equation}
where $f_m$ is the modulation frequency of the sine and cosine waves.
The generated waveform is shown in \figpanel{fig:Sawtooth_sim}{a}.
In a Bloch-sphere picture, this rotates the Bloch vector in the x-y plane at a constant rate $f_m$.
Through this modulation, the carrier signal can be detuned, allowing control over efficiency by adjusting $f_m$ as demonstrated in \figpanel{fig:Sawtooth_sim}{d}, similar to \figpanel{fig:swp_N}{f}.
The simulated output voltages and excited-state population for sawtooth pulses are also shown in \figref{fig:Sawtooth_sim}.

To ensure compliance with the equipment bandwidth, it is crucial to minimize discontinuities in $\Pi(t)$ during the onset of the rising pulse.
Discontinuities like $0-\theta-0-\theta$ generate higher-order harmonics in IQ voltages that may be filtered out by the equipment bandwidth.
In linear modulation [\eqref{Eq:phaseMod_linear}], $\Pi(t)$ remains continuous, and the spectrum of the rising pulse is centered at $f_m$ within the equipment bandwidth, effectively avoiding filtering.

\bibliography{PhaseShapingRefs}

\end{document}